# Transparent planar solar absorber for winter thermal management

Muhammad Saad Asad[†,*]

Muhammad Zulfiker Alam[†]

[†]Department of Electrical Engineering, Queen's University, Kingston, Canada

[*]Corresponding author.    Email: 19msa4@queensu.ca

## Abstract

Indoor heating during winters accounts for a significant portion of energy consumed by buildings in regions of cold climate. Development of transparent coatings for windows that efficiently harvest solar energy can play a major role in reducing energy consumption and fuel costs incurred for winter heating. In recent years, there has been a great research effort towards designing transparent solar absorber coatings using nanophotonic structures. The potential of coatings based on planar multilayer structures, however, has received very little attention. In this work we investigate the performance of planar multilayer thin films using low cost materials for design of transparent solar absorber window coatings. Our study led to the proposal of two planar multilayer designs. Simulation results predict that an increase in surface temperature by 21 K and 25 K, while maintaining mean visible transmittance of over 60% is possible using these designs. These results illustrate the great promise planar multilayer structures hold for winter thermal management of buildings.

## 1. Introduction

Buildings are responsible for more than 30% of the energy consumption worldwide [1,2] and 19% of the global green house gas emission [3]. Indoor heating in cold climates is a significant contributor to the energy demand of buildings. The use of fossil fuel for heating not only negatively affects the environment, but the high cost of fossil fuels also results in widespread energy poverty [4]. There is an urgent need to develop simple and inexpensive methods of passive heating to lower building energy usage and costs incurred for winter heating. Current methods of passive heating including double-glazed window, Trombe wall and solarium have several limitations. These schemes often require significant investment and cannot be easily retrofitted to existing buildings [5]. In recent years, the rapid progress of nanophotonics has provided us greatly improved control on light-matter interaction allowing new ways of harnessing solar energy for passive heating. This has led to the development of transparent solar absorber coatings based on metamaterial and metasurface architectures for passive winter thermal management ranging from applications of indoor heating [6-9] to de-icing and antifogging applications [10-12].

Windows are an important part of most buildings and are often responsible for more than 25% of heat loss in a typical building [13]. Unlike walls, windows need to be transparent to visible light, making it challenging to avoid heat loss through them. Furthermore, in addition to being felt cold while touched, a cold window will result in a draft, causing discomfort anywhere in the room. Thermal comfort is not only influenced by the temperature of the indoor air but also by the temperature of indoor surfaces including walls, windows and floors which contribute to the mean radiant temperature (MRT) of an indoor space. During cold days, increasing the MRT by 3 °C can allow the indoor air temperature to be decreased by 2 °C without affecting indoor thermal comfort [6,14,15]. More than 50% of the solar energy lies in the near-infrared (NIR) region, which does not contribute to illumination [13]. By designing specialized window coatings that effectively absorb NIR solar radiation to raise the surface temperature one can, in turn, raise the MRT of an indoor space. Such technology will allow the possibility of lowering the indoor air temperature without affecting thermal comfort and hence reduce both cost and energy demand for indoor heating.

Optical metamaterial and metasurface have proven to be a remarkably successful technology for implementation of transparent solar absorber coatings. 8 K rise in surface temperature of a glass substrate while maintaining 75% overall visible transmittance was reported using a metamaterial design based on nickel nanoantennas [6]. Application of hybrid plasmonic metasurface coating based on gold nanoparticles embedded in a $TiO_2$ matrix for antifogging application was investigated in [12] which achieved a mean visible transmittance of 36%. An asymmetric metasurface architecture based on periodic titanium/aluminum oxide/copper arrays for transparent solar absorber was reported in [8]. Numerical analysis predicted a mean visible transmittance above 60% and mean absorption of 45% for this design. Among more recent works, a planar linear gold and silver nanoparticle assembly coated on a glass substrate demonstrated temperature increase by 9.8 K while visible transparency was larger than 65% [9]. Another recent work demonstrated 38 K rise in surface temperature using a coating based on Cesium-doped tungsten oxide nanoparticles for antifogging applications [10].

Previous works on passive heating focused on either periodic structures or nanoparticles. Metal/insulator/metal (MIM) is an important class of nanophotonic structures that has been investigated for many applications including broadband solar absorption [16-20] and selective visible and NIR filters [21] due to its advantages e.g. ability to enhance light-matter interaction and ease of fabrication. However, the potential of planar MIM architecture for development of transparent solar energy absorber coatings has received very little attention. In this work, we carried out a systematic analysis of the potential of planar MIM structure for such an application with a focus on designs that can be implemented using low cost materials.

The rest of the paper is organized as follows. In Section 2 we reviewed the design geometry investigated in this work and methodology used for optical analysis. In Section 3 we carried out a detailed parametric study using a combination of transfer matrix method and particle swarm optimization (PSO) to determine the optimum choices of materials and device dimensions. We focused on designs which can be fabricated using low cost materials and inexpensive and scalable fabrication process. We found that TiN/SiO$_2$/TiN configuration provides a combination of good transmission in the visible wavelength range and good overall absorption. Further analysis revealed that addition of two thin TiO$_2$ layers simultaneously enhances maximum transmittance in the visible region and mean absorption. In Section 4 we carried out thermal simulation for these two optimized designs and found that under AM 1.5 solar illumination these designs are capable of increasing the steady-state temperature of a glass surface by 21 K and 25 K respectively. These results are significantly better than many previously reported nanophotonic structures for passive heating of windows, which requires the use of noble metals or complex fabrication process steps. Section 5 concludes the paper with comparison of our designs with available alternatives, and suggestions about potential applications of our designs.

## 2. Overview of the design process

The purpose of this work is to develop a transparent solar absorber that can be fabricated using low cost materials with simple, inexpensive, and scalable fabrication process. To achieve this, we chose to base our initial design on the MIM architecture shown in Fig. 1 because it can be fabricated using simple thin film deposition process and does not require any patterning step. We attempted to find designs, which show high visible light transparency (400-700 nm) and high NIR absorption (700-2500 nm). As described in [19,20], even for a simple MIM structure, the light-matter interaction can be quite complex. Therefore, utilizing a suitable optimization algorithm is required for finding the optimum solution. We use a combination of PSO and transfer matrix formulation to optimize our designs. Application of transfer matrix allows fast and accurate determination of transmission, reflection, and absorption of planar multilayer stacks [22,23]. In the following discussion we will briefly review the transfer matrix method. This discussion closely follows the description reported in [22]

Fig. 2 shows a multilayer structure consisting of $N$ layers of finite thicknesses sandwiched between a cladding (layer 0) and a substrate (layer $N+1$). The thickness of the $ith$ layer is given by $d_i$ and its complex refractive index is $\hat{n}_i = n_i + jk_i$.

$\theta_i$ denotes the angle at which light propagates in medium $i$ with respect to the normal at the interface of medium $i-1$ and $i$. The transfer matrix relating the electric field entering and leaving the stack is given by

$$M = \begin{pmatrix} M_{11} & M_{12} \\ M_{21} & M_{22} \end{pmatrix} = D_{0,1} P_1 D_{1,2} P_2 D_{2,3} P_3 \dots P_N D_{N,N+1} \qquad (1)$$

Here $D_{i,i+1}$ is the matrix relating the forward and backward propagating electric field quantities at the two sides of the interface between layer $i$ and layer $i+1$ ($i = 0,1,..,N$).

$$D_{i,i+1} = \frac{1}{t_{i,i+1}} \begin{bmatrix} 1 & r_{i,i+1} \\ r_{i,i+1} & 1 \end{bmatrix} \qquad (2)$$

Here $r_{i,i+1}$ and $t_{i,i+1}$ are Fresnel reflection and transmission coefficients between layer $i$ and $i+1$. For a TE or s polarized incident wave, they are given by

$$r_{i,i+1} = \frac{\hat{n}_i \cos\theta_i - \hat{n}_{i+1} \cos\theta_{i+1}}{\hat{n}_i \cos\theta_i + \hat{n}_{i+1} \cos\theta_{i+1}} \tag{3}$$

$$t_{i,i+1} = \frac{2\hat{n}_i \cos\theta_i}{\hat{n}_i \cos\theta_i + \hat{n}_{i+1} \cos\theta_{i+1}} \tag{4}$$

For a TM or p polarized incident wave, these quantities are given by

$$r_{i,i+1} = \frac{\hat{n}_i \cos\theta_{i+1} - \hat{n}_{i+1} \cos\theta_i}{\hat{n}_i \cos\theta_{i+1} + \hat{n}_{i+1} \cos\theta_i} \tag{5}$$

$$t_{i,i+1} = \frac{2\hat{n}_i \cos\theta_i}{\hat{n}_i \cos\theta_{i+1} + \hat{n}_{i+1} \cos\theta_i} \tag{6}$$

The field quantities at the top and bottom interfaces of the $ith$ layer are related by the propagation matrix $P_i$ ($i = 1, \ldots, N$) which is given by

$$P_i = \begin{bmatrix} e^{-j\varphi_i} & 0 \\ 0 & e^{+j\varphi_i} \end{bmatrix} \tag{7}$$

Where $\varphi_i$ is the phase shift experienced by the wave when it propagates a distance $d_i$ through the $ith$ layer, and has the following expression

$$\varphi_i = \frac{2\pi}{\lambda} \hat{n}_i d_i \cos\theta_i \tag{8}$$

Once the transfer matrix $M$ is calculated, the reflection $R$ and transmission $T$ of the multilayer stack can be found by using the following expressions,

$$R = \left|\frac{M_{21}}{M_{11}}\right|^2 \tag{9}$$

$$T = \frac{n_{N+1}\cos\theta_{N+1}}{n_0\cos\theta_0}\left|\frac{1}{M_{11}}\right|^2 \tag{10}$$

Where $n_{N+1}$ is the real part of the refractive index of the substrate and $n_0$ is the real part of the refractive index of the cladding.

The absorption of the multilayer stack can be determined using the relation

$$A(\lambda) = 1 - R(\lambda) - T(\lambda) \tag{11}$$

To optimize our designs, we chose to use PSO due to its ability to find an optimum solution from a very large solution space without the need to make assumptions about the nature of the optimization problem [24-28]. In PSO, the optimization process starts with several trial solutions, which are called particles. After each iteration, each particle updates its local optimal solution found near its vicinity as well as the global optimal solution found because of the movement of all particles through the search space. We carried out the optimization in two steps. In step 1, we investigated the simple MIM architecture sandwiched between $SiO_2$ substrate and cladding. This analysis allowed us to identify the optimal choice of materials and layer thicknesses for the MIM design. In Step 2, we extended this analysis and added additional $TiO_2$ layers on top and bottom of the optimized MIM structure and re-optimized the layer thicknesses.

### 3. Optimization of MIM structure for passive heating

*3.1 Determination of optimal material choice and layer thicknesses*

To determine the optimal material choice and layer thickness of the MIM design, we first investigated different plasmonic materials for the top and bottom layers. The materials investigated included Al, Cu, Ni, TiN and Au. Al has been extensively studied due to its strong plasmon response in the visible region [29,30]. Cu and Ni are low cost metals and have been studied by other groups for transparent thermal management applications [6-8,31]. Transition metal nitride TiN is considered a good candidate due to its low cost and higher mechanical and thermal stability than noble metals [32,33]. We included Au in our investigation to determine the possible compromise in performance that we may have to accept when we use cheaper plasmonic materials. Optical properties of Cu, Al and Au were taken from [34], and those for Ni were taken [35]. For TiN, the refractive index data was obtained from [36,37]. The dielectric spacer region separating the metallic layers for this analysis is assumed to be $SiO_2$. Optical properties of $SiO_2$ were taken from [35]. Effect of the choosing other spacer materials will be analyzed later in the paper.

As stated in Section 2, since our designs are based on planar multilayer MIM stacks, transfer matrix formulation together with PSO was used to simultaneously maximize both visible transmittance and NIR absorption. The transfer matrix method was implemented using MATLAB. The accuracy of this code was verified by comparing the results obtained from this code with those obtained from Lumerical STACK from Ansys. The optimization procedure was implemented using the PSO toolbox provided by MATLAB.

For the optimization procedure, we aimed to maximize two objective functions: the mean visible transmittance and the mean NIR absorption. In order to simultaneously maximize both quantities we utilized the conventional weighted aggregation (CWA) approach for multi-objective optimization [27,28]. In this method, all objective functions are summed to form one single objective function that the optimization algorithm aims to either maximize or minimize. Each objective function is associated with a non-negative weight that remains fixed during the optimization. We used the following figure of merit (FOM) to optimize our planar multilayer designs.

$$FOM = w_1 mean(T_{vis}) + w_2 mean(A_{NIR}) \qquad (12)$$

Where $T_{vis}$ is the visible transmittance, $A_{NIR}$ is the absorption in the NIR region and $w_1$ and $w_2$ are non-negative weights such that

$$w_1 + w_2 = 1 \qquad (13)$$

For the optimization we used 15 particles per iteration and carried out 30 iterations, running a total of 450 simulations for each material configuration. The layer thicknesses were set as the design parameter. The upper and lower limits set for the thickness parameter were 2 nm and 50 nm for the top and bottom layers and 2 nm and 500 nm for the middle $SiO_2$ dielectric spacer.

The layer thicknesses and the optical performance of the optimized designs are summarized in Table 1. The last column of the table gives the solar power absorbed by each design under AM 1.5 illumination. This data was obtained by first multiplying the normalized absorption profile with the solar spectrum (AM 1.5), followed by integrating the resulting spectra over the 280 nm to 2500 nm wavelength range to calculate the absorbed power density. Fig. 3 shows transmittance and absorption profiles obtained for these optimized designs. Au has been extensively used in the past for transparent passive heating applications [6,8,9,11,12]. However, our study reveals that cheaper alternatives perform much better for these applications in case of the MIM geometry. Both the TiN and Ni-based designs significantly outperform the other designs in terms of mean transmittance in the visible region and absorbed solar power density. Although the former provides lower solar power absorption, the transmission is 15.6% higher for the former compared to the latter. Therefore, we chose the TiN-based design for further investigation. Although the optical response of this design was presented in Fig. 3 with those for other designs, we reproduce the optical response for the TiN-based design in Fig. 4 for ease of visualization.

The analysis presented so far assumed that the spacer material separating the metallic layers is $SiO_2$. We now consider two alternate choices of spacer material: silicon nitride (SiN) and aluminum oxide ($Al_2O_3$). Refractive index data for SiN and $Al_2O_3$ are taken from [38] and [35] respectively. We reoptimized our designs for these new choices of spacer materials. Table 2 summarizes the optimized designs and key performance results for the three choices of spacer material. The optical response for these three designs is presented in Fig. 5. Among the three choices for spacer materials we considered, the SiN design shows the lowest mean visible transmittance and least solar power absorption. Although the $Al_2O_3$ design absorbs 20.6 W/m$^2$ more solar power compared to the $SiO_2$ design, the latter design shows a significantly higher mean visible transmittance. Therefore, we chose $SiO_2$ to be the spacer medium for the rest of our paper.

*3.2 Effect of adding additional $TiO_2$ layers to the $TiN/SiO_2/TiN$ MIM design.*

Our choice of MIM geometry for passive solar heating was motivated by previous reports on the applications of this geometry for the design of perfect absorbers [16-20]. The inverse of this geometry (insulator/metal/insulator or IMI), on the other hand has been used for enhancing transmission [31,39-42]. We now investigate a combination of MIM and IMI structures to see if such a combination can provide superior performance than what is achievable from MIM alone. As shown in Fig. 6a, the new structure is achieved by adding $TiO_2$ layers on the two sides of the $TiN/SiO_2/TiN$ structure analyzed in Section 3. $TiO_2$ has a higher refractive index than $SiO_2$ and has high mechanical strength, which makes it a suitable choice for our design [31,41,42].

As done in Section 3.1, transfer matrix formulation in conjunction with PSO is utilized to optimize the layer thicknesses. The upper and lower limits set for the thickness parameter were 2 nm and 50 nm for the TiN layer and 2 nm and 500 nm for the dielectric layers. Addition of two $TiO_2$ layers increased the complexity of optimization, therefore the total number of particles used per iteration was increased to 30. 40 iterations were carried out, running a total of 1200 simulations. The optimized design and its optical response are shown in Fig. 6a and 6b respectively. A comparison of the optical response before and after addition of the $TiO_2$ layers to the $TiN/SiO_2/TiN$ design is presented in Table 3. We designate the $TiN/SiO_2/TiN$ and the $TiO_2/TiN/SiO_2/TiN/TiO_2$ designs as Design 1 and Design 2 respectively. The addition of $TiO_2$ layers not only enhances solar absorption compared to Design 1 but also enhances the maximum visible transmittance, which reaches values as high as 80%.

## 4. Thermal Simulations

In Section 3 we have identified two designs which provide good transmission in the visible region, while absorbing a significant amount of solar energy. In this section, we examine the temperature rise achievable when the proposed structures are coated on a glass substrate. The results presented here are obtained using a combination of Lumerical FDTD and Lumerical HEAT Solver from Ansys. A schematic of the simulation setup for Design 1 is shown in Fig. 7. We first carried out

2D FDTD simulations to determine the spatial absorption under normal illumination which is then multiplied by the solar spectrum (AM 1.5) and integrated over the 280 nm to 2500 nm wavelength range to determine the spatial distribution of absorbed power in the xy plane over the area bounded by the dashed yellow lines in Fig. 7. This absorption data was then exported to HEAT. For thermal simulation in HEAT, the x-span was set to 0.4 µm similar to the one used for FDTD simulation, and the planar multilayer structure was placed on an 0.4 mm thick $SiO_2$ substrate. A 100 µm $SiO_2$ coating was applied to the top of the MIM structure to act as a cladding. The ambient temperature of the simulation was kept at 300 K. Convective boundary conditions were applied to the top and bottom of the simulation region to simulate heat dissipation from the device and the heat transfer coefficient was set to 10 $Wm^{-2}K^{-1}$ which is a common value used for air. A similar simulation setup was also used for Design 2. The thermal simulations predicted that under AM 1.5 solar illumination the steady state temperature of the glass substrate will rise by 21 K and 25 K respectively. As discussed in more detail in the next section, these designs significantly outperform many previous transparent solar absorber designs.

5. **Conclusion**

Application of nanophotonics for passive heating of buildings can play a significant role in reduction of global greenhouse gas emission and energy poverty. In this work we studied the potential of planar multilayer structures using low cost plasmonic materials for this application. Our study led to the proposal of two planar multilayer designs. Design 1 is a $TiN/SiO_2/TiN$ MIM design, and Design 2 is a $TiO_2/TiN/SiO_2/TiN/TiO_2$ design, which is a combination of MIM and IMI architectures. Both designs can be implemented using low cost materials and simple, inexpensive, and scalable fabrication process. The designs proposed outperform alternatives that require more expensive materials or complicated fabrication process steps. For example, both Design 1 and 2 are predicted to provide temperature rise which is more than double than that reported in [6]. Fabrication of the structure reported in [6] requires multiple steps including deposition, etching and lift off. In contrast, fabrication of our proposed designs will require only deposition of multiple layers of thin films. Another very recent work proposed an asymmetric metasurface design which predicted a mean visible transmittance of above 60% and mean absorption of 45% [8]. Similar to the design reported in [6], it also requires metal deposition, etching and lift off for its fabrication. The mean absorption for Design 1 (calculated but not shown in Table 1) is 3% lower than that reported in [8] but the mean visible transmittance is 8% higher. Design 2 exhibits 7% higher mean absorption than that reported in [8] while demonstrating similar mean transmittance.

The proposed nanophotonic designs for passive heating can be adapted to windows for buildings in a number of ways. They can be deposited directly on a glass window or on a transparent polymer coating, which can be attached to windows to achieve passive heating only during winter. It is also possible to integrate these multiplayer structures with commercially available low-emissivity (low-E) window coating technology [13]. Such a window will be able to simultaneously absorb solar energy for passive heating and prevent escape of heat from indoor by reflection. The benefit of such a technology for reduction in energy consumption would be significant for modern residential and commercial buildings, where windows form a large part of building surface. We believe the low cost, ease of fabrication and ability to provide large temperature increase make our designs attractive choices for passive heating of buildings and will encourage further exploration of planar multilayer structures for passive heating applications.

## Acknowledgement

We acknowledge the support of the Natural Sciences and Engineering Research Council of Canada (NSERC).

Academic Press, Burlington, 1997: pp. 293–311.

**Figure Captions**

FIG 1    Schematic of the MIM design investigated for passive heating.

FIG 2    Planar multilayer structure.

FIG 3    (a) Transmittance 'T' and (b) absorption 'A' spectra for the optimized designs for different choices of top and bottom materials.

FIG 4    (a) Schematic and (b) optical response of the optimized TiN/SiO$_2$/TiN MIM design.

FIG 5    (a) Transmittance 'T' and (b) absorption 'A' spectra for the optimized designs for various choices of spacer materials. The plasmonic material is TiN for all the designs.

FIG 6    (a) Proposed TiO$_2$/TiN/SiO$_2$/TiN/TiO$_2$ design based on insulator/MIM/insulator architecture and (b) optical response the optimized design.

FIG 7    Thermal Simulation setup for Design 1.

**Table 1**

Summary of layer thicknesses and optical performance for the optimized MIM design for various choices of metals. The spacer is $SiO_2$.

| $t1$ (nm) | $t2$ (nm) | $t3$ (nm) | Metal/Metal nitride | Mean transmittance in visible (400-700 nm) | Maximum transmittance in visible and its location (400-700 nm) | Absorbed solar power density (W/m$^2$) |
|---|---|---|---|---|---|---|
| 2 | 402 | 3 | Al | 0.523 | 0.727 (441 nm) | 369.4 |
| 6 | 404 | 16 | Au | 0.498 | 0.711 (700 nm) | 179.1 |
| 4 | 398 | 10 | Cu | 0.543 | 0.763 (700 nm) | 253.9 |
| 2 | 150 | 6 | Ni | 0.530 | 0.55 (459 nm) | 490.0 |
| 5 | 155 | 5 | TiN | 0.686 | 0.725 (454 nm) | 421.3 |

**Table 2**

Summary of layer thicknesses and optical performance for the optimized MIM design for various choices of spacer materials. The metallic layers are TiN for all the designs.

| $t1$ (nm) | $t2$ (nm) | $t3$ (nm) | Dielectric | Mean transmittance in visible (400-700 nm) | Maximum transmittance in visible and its location (400-700 nm) | Absorbed solar power density (W/m$^2$) |
|---|---|---|---|---|---|---|
| 5 | 155 | 5 | SiO$_2$ | 0.686 | 0.725 (454 nm) | 421.3 |
| 4 | 114 | 8 | Al$_2$O$_3$ | 0.638 | 0.683 (445 nm) | 441.9 |
| 3 | 95 | 9 | SiN | 0.623 | 0.678 (434 nm) | 420.7 |

**Table 3**

Summary of optical and thermal performance results Design 1 and Design 2.

| Multilayer design | Mean transmittance in visible (400-700 nm) | Maximum transmittance in visible and its location (400-700 nm) | Absorbed solar power density (W/m$^2$) | Simulated temperature increase (K) |
|---|---|---|---|---|
| Design 1 | 0.686 | 0.725 (454 nm) | 421.3 | 21 |
| Design 2 | 0.605 | 0.803 (459 nm) | 508.5 | 25 |

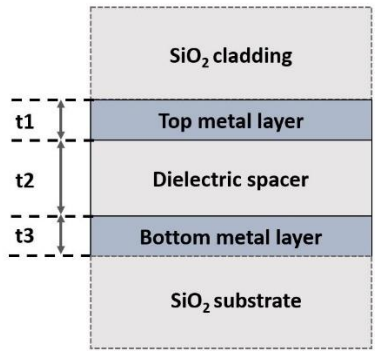

FIG. 1 Asad & Alam 2021

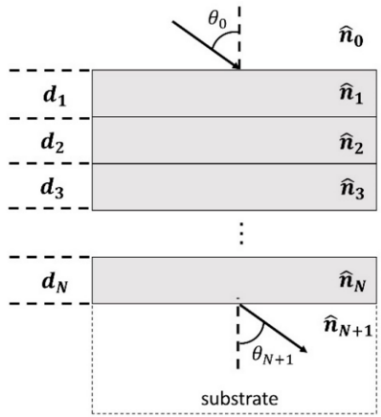

FIG. 2 Asad & Alam 2021

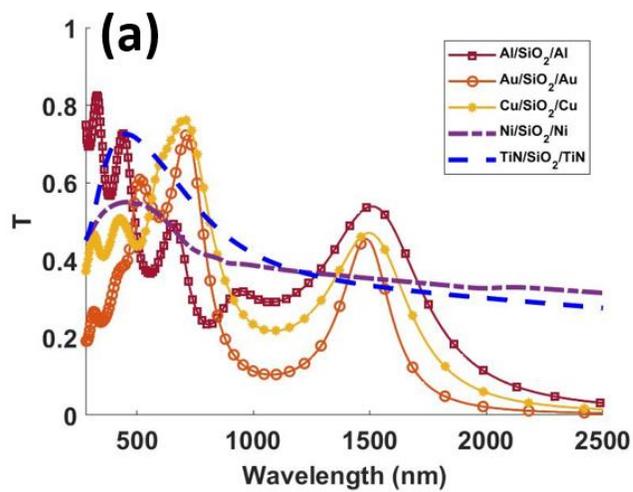
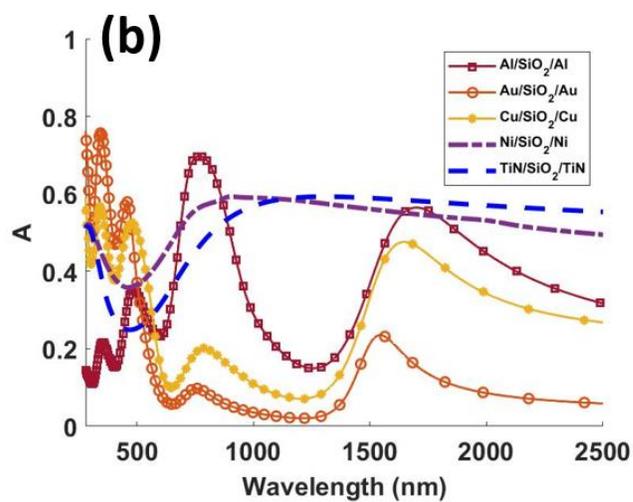

FIG. 3 Asad & Alam 2021

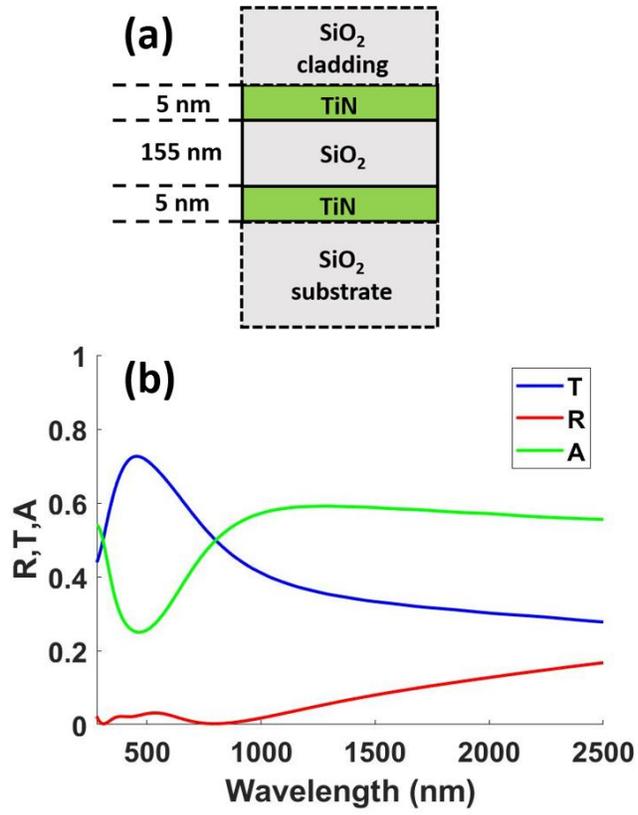

FIG. 4 Asad & Alam 2021

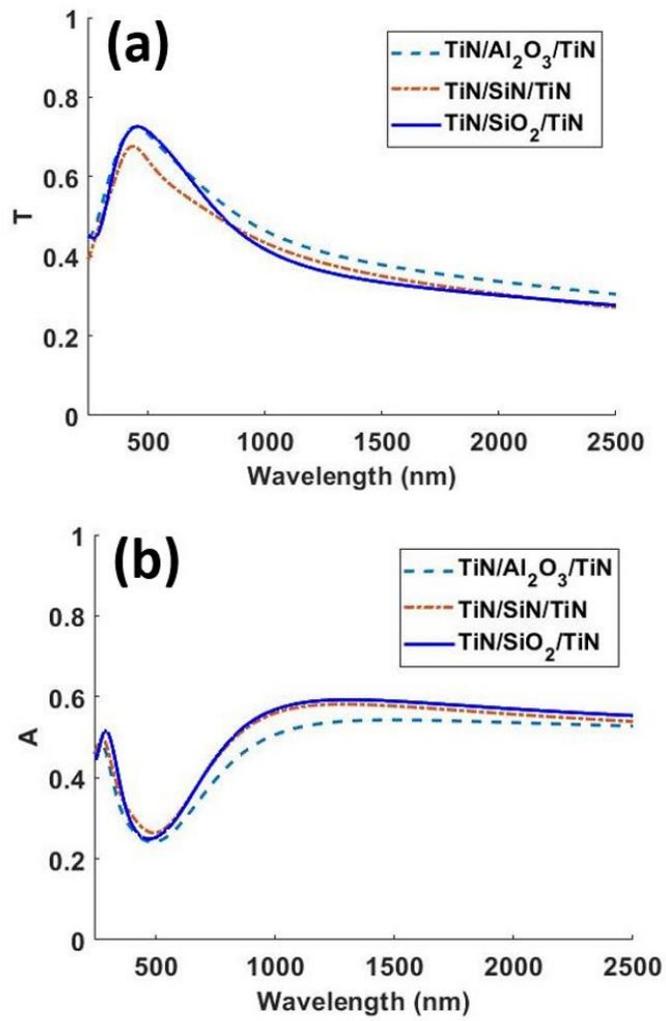

FIG. 5 Asad & Alam 2021

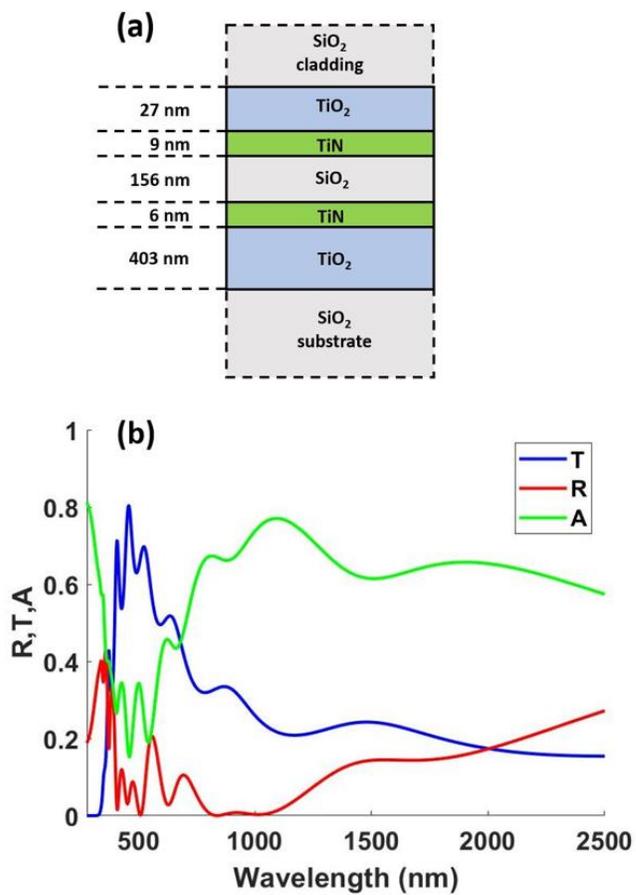

FIG. 6 Asad & Alam 2021

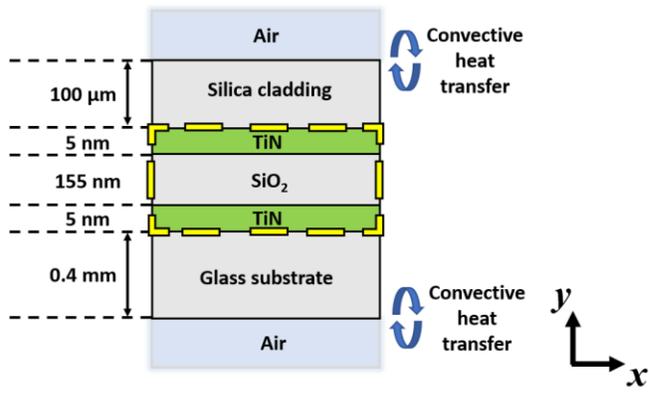

FIG. 7 Asad & Alam 2021